\documentclass[aps,prx,twocolumn,superscriptaddress,export,longbibliography]{revtex4-2}

\usepackage{amsmath}
\usepackage{bm}
\usepackage[dvipsnames]{xcolor}
\usepackage{hyperref}
\hypersetup{
	colorlinks=true,
	linkcolor=blue,
        citecolor=blue,
	urlcolor=blue,}
 \usepackage{graphicx}
 \usepackage{tikz}

\newcommand{\grad}{\nabla}
\newcommand{\goesto}{\rightarrow}
\newcommand{\eff}{\text{eff}}
\renewcommand{\L}{\mathcal{L}}

\newcommand\invisiblesection[1]{%
  \refstepcounter{section}%
  \addcontentsline{toc}{section}{\protect\numberline{\thesection}#1}%
  \sectionmark{#1}}

\begin{document}

\title{Fracton superfluid hydrodynamics}
\author{Charles Stahl}
\affiliation{Department of Physics and Center for Theory of Quantum Matter, University of Colorado, Boulder, CO 80309, USA}
\author{Marvin Qi}
\affiliation{Department of Physics and Center for Theory of Quantum Matter, University of Colorado, Boulder, CO 80309, USA}
\author{Paolo Glorioso}
\affiliation{Department of Physics, Stanford University, Stanford CA 94305, USA}
\author{Andrew Lucas}
\affiliation{Department of Physics and Center for Theory of Quantum Matter, University of Colorado, Boulder, CO 80309, USA}
\author{Rahul Nandkishore}
\affiliation{Department of Physics and Center for Theory of Quantum Matter, University of Colorado, Boulder, CO 80309, USA}

\begin{abstract}

We examine the hydrodynamics of systems with spontaneously broken multipolar symmetries using a systematic effective field theory. We focus on the simplest non-trivial setting: a system with charge and dipole symmetry, but without momentum conservation. When no symmetries are broken, our formalism reproduces the quartic subdiffusion ($\omega \sim -i k^4$) characteristic of `fracton hydrodynamics' with conserved dipole moment. Our formalism also captures spontaneous breaking of charge and/or dipole symmetry. When charge symmetry is spontaneously broken, the hydrodynamic modes are quadratically propagating and quartically relaxing ($\omega \sim \pm k^2 - ik^4$). When the dipole symmetry is spontaneously broken but the charge symmetry is preserved, then we find quadratically relaxing (diffusive) transverse modes, plus another mode which depending on parameters may be either purely diffusive ($\omega \sim -i k^2$) or quadratically propagating and quadratically relaxing ($\omega \sim \pm k^2 -i k^2$). Our work provides concrete predictions that may be tested in near-term cold atom experiments, and also lays out a general framework that may be applied to study systems with spontaneously broken multipolar symmetries. 

\end{abstract}
	
\date{\today}
	
\maketitle
\invisiblesection{Introduction}
\textit{Introduction.}---
Multipolar symmetries are exciting widespread interest in modern condensed matter physics, quantum information, and quantum dynamics. They connect to exotic `fracton' phases of quantum matter \cite{PretkoSubdimensional, GromovMultipoleAlgebra} and can provide a new route to ergodicity breaking \cite{PPN, Khemani2020, SalaFragmentation2020}. Multipolar symmetries can display partial or complete spontaneous symmetry breaking (SSB), through which they can stabilize new kinds of phases \cite{Lake2022}. Of particular interest, the approach to equilibrium in systems with multipolar symmetries is described by hydrodynamics in an infinite family of non-standard universality classes \cite{fractonhydro, ivn}, collectively termed `fracton hydrodynamics.' This `fracton hydrodynamics'---which has been realized in ultracold atoms \cite{GuardadoSanchez2020}---provides an exciting new frontier, the exploration of which has become an important topic of research in its own right \cite{gloriosobreakdown, RichterPal, IaconisMultipole, nonabelian, grosvenor, osborne, Feldmeier, SalaModulated, HartQuasiconservation, fractonmhd,Guo:2022ixk,gloriosoSSB}. 

The {\it thermodynamics} of SSB of multipolar symmetries has been discussed in \cite{Stahl2022, Kapustin}, where analogs of the Mermin-Wagner and Imry-Ma theorems were established. Different patterns of SSB, either of the entire multipole group or of its subgroups, correspond to condensing either monopole charges or higher multipole charges. We will call such SSB phases `fracton superfluids.'
The hydrodynamics of conventional superfluids is a well studied subject \cite{superfluidhydro}---here, the Goldstone boson of the broken symmetry becomes a hydrodynamic mode. However, given the surprises attendant in the  hydrodynamics of systems with unbroken multipolar symmetries, one might anticipate new features in hydrodynamics of fracton superfluids. By analogy, we may term the hydrodynamics of such generalized superfluids `fracton superfluid hydrodynamics.' Previous literature has studied fracton superfluids at zero temperature~\cite{Yuan2020,Lake2022}. 
Other work has shown that multipolar symmetries and translation symmetry together lead to exotic hydrodynamics in which one symmetry \emph{must} be spontaneously broken~\cite{gloriosoSSB}, by certain definitions of SSB; see also \cite{Jensen:2022iww}. 

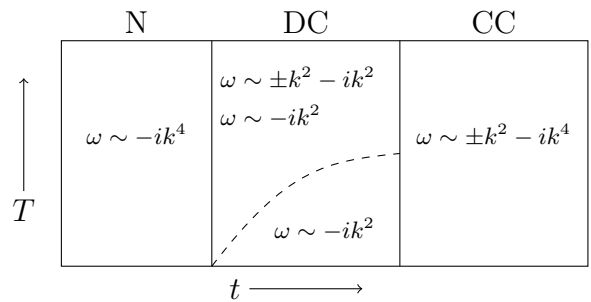
\begin{figure}
    \centering
\begin{tikzpicture}
\draw (0,0) rectangle (7,3);
\draw (2,0) -- (2,3);
\draw (4.5,0) -- (4.5,3);
\draw[dashed] (2,0) .. controls (3,1.3) and (3.5,1.4) .. (4.5,1.5);
\draw[left,->] (2.5,-.3) node {\large $t$} -- (4,-.3);
\draw[below,->] (-.5,1) node {\large $T$} -- (-.5,2.5);
\node at (1,1.75) { $\omega \sim -ik^4$};
\node at (3.5,.5) { $\omega \sim -ik^2$};
\node[right] at (2,2.5) { $\omega \sim \pm k^2 -ik^2$};
\node[right] at (2,2) { $\omega \sim -ik^2$};
\node at (5.75,1.75) { $\omega \sim \pm k^2 - ik^4$};
\node at (1,3.25) {\large N};
\node at (3.25,3.25) {\large DC};
\node at (5.75,3.25) {\large CC};
\end{tikzpicture}
    \caption{A rough organization of the three phases we find. The axes are temperature $T$ and a generic hopping coefficient $t$ that condenses dipole charges and monopole charges successively. The normal phase (N) displays subdiffusion while the charge condensate (CC) displays quadratically-propagating modes. The dipole condensate (DC) phase displays a crossover between only diffusive modes at small $T$ and coexisting diffusive and propagating modes at large $T$.}
    \label{fig:phasediagram}
\end{figure}

In this Letter, we develop the theory of fracton superfluid hydrodynamics at nonzero temperatures. We do so in the simplest possible setting---a system with only charge and dipole symmetry, leaving generalization to arbitrary multipole groups and/or momentum conservation to future work. 
For such charge and dipole conserving systems, we develop a systematic effective field theory description which yields three phases, roughly organized as in Fig.~\ref{fig:phasediagram}. One phase corresponds to the `fracton hydrodynamics' of~\cite{fractonhydro} with no symmetries broken. Another phase, with the symmetry fully broken, has been called the `fractonic superfluid'~\cite{Yuan2020} or the `Bose-Einstein insulator'~\cite{Lake2022}. We will call it the `charge condensate'. Finally, the `dipole condensate', with the dipole symmetry spontaneously broken and the monopole symmetry unbroken, exhibits new hydrodynamics. The two condensate phases are both fracton superfluids.

\invisiblesection{Effective action}
\textit{Effective action.}---
We will use the recently-developed hydrodynamic effective field theory~\cite{Crossley:2015evo,Haehl:2015foa,Jensen:2017kzi,Glorioso:2017fpd,liu2018lectures} to explore our hydrodynamic phases.
To build the effective action for a system with charge and dipole conservation we will use the phase fields $\phi$ and $\psi_i$. The action should be invariant under global dipole transformations parametrized by arbitrary constants $b$ and $c_i$, under which the fields transform as 
\begin{align}
\phi &\goesto \phi + b + x^k c_k,\nonumber\\
\psi_i &\goesto \psi_i + c_i.
\end{align}
The invariant objects are $\partial_t\phi$, $\partial_t\psi_i$, $\grad_i\phi-\psi_i$, $\grad_i\psi_j$, and $\grad_i\grad_j\phi$. Here $\phi$ is the `monopole' field and $\psi_i$ is the `dipole' field, and the combination $\grad_i\phi-\psi_i$ indicates that the motion of a monopole charge involves absorption (or emission) of a dipole. We notice that $\grad_i\grad_j\phi=\grad_i(\grad_j\phi-\psi_j)+\grad_i\psi_j$, demonstrating that $\grad_i\grad_j \phi$ is a redundant degree of freedom. 

Our most general Lagrangian is then
\begin{align}
\L = \L(\partial_t\phi, \partial_t\psi_i, \grad_i\phi-\psi_i, \grad_i\psi_j). \label{eqn:singlecontour}
\end{align}
From this, we can derive two Noether-like equations:
\begin{align}
0 &= \partial_t \rho + \grad_i J_i, \label{eqn:monopole}\\
0 &= \partial_t \rho_i + J_i - \grad_j J_{ij} \label{eqn:dipole}
\end{align}  
where we have defined
\begin{align}
\rho &\equiv \frac{\partial \L}{\partial (\partial_t \phi)} & J_i &\equiv \frac{\partial \L}{\partial(\grad_i\phi)} = -\frac{\partial \L}{\partial \psi_i}\nonumber\\
\rho_i &\equiv \frac{\partial \L }{\partial (\partial_t \psi_i)} & J_{ij} &\equiv -\frac{\partial \L}{\partial(\grad_j \psi_i)}
\end{align}
as the densities and currents. We will call~\eqref{eqn:monopole} the monopole continuity equation, and~\eqref{eqn:dipole} the dipole continuity equation.
To recognize~\eqref{eqn:dipole} as dipole conservation, we can define the total dipole moment $d_i \equiv x_i\rho + \rho_i$ so that it obeys a continuity equation
\begin{align}
0 = \partial_t d_i + \grad_j J_{ij}^{(d)},
\end{align}
where $J_{ij}^{(d)} = x_iJ_j - J_{ij}$.

In order to turn this into a hydrodynamic EFT, following~\cite{liu2018lectures}, we must put the action on a doubled Schwinger-Keldysh contour, and define forward-propagating fields $\phi_1$ and $\psi_{i1}$ and backward-propagating fields $\phi_2$ and $\psi_{i2}$ on the two contours. In the hydrodynamic limit the forward and backward fields are close to equal so it easier to work with the ``classical fields" $\phi = (\phi_1+\phi_2)/2$, $\psi_i = (\psi_{i1}+\psi_{i2})/2$ and the ``noise fields" $\Phi = \phi_1 - \phi_2$, $\Psi_i = \psi_{i1}-\psi_{i2}$.

The full hydrodynamic Lagrangian will have the form 
\begin{align}
\L_\eff = \rho \partial_t \Phi + \rho_i \partial_t  \Psi_i + J_i (\grad_i \Phi- \Psi_i) - J_{ij} \grad_j\Psi_i, \label{eqn:hydro}
\end{align}
where the densities and currents may depend on $\phi$, $\psi_i$, $\Phi$, and $\Psi_i$. 
We have no terms with $\grad_i\grad_j \Phi$ for the same reason we have no terms involving $\grad_i\grad_j \phi$ in~\eqref{eqn:singlecontour}: such terms can be converted into the terms already present in~\eqref{eqn:hydro}. For details, see the Appendix.

The action $I[\phi, \Phi, \psi_i, \Psi_i] = \int d^3x\, dt\, \L_\eff$ must be symmetric under
\begin{align}
\phi &\goesto \phi + b + x^k c_k, & \psi_i &\goesto \psi_i + c_i, \nonumber\\
\Phi &\goesto \Phi + b' + x^k c_k', & \Psi_i &\goesto \Psi_i + c'_i,
\end{align}
which are independent dipole transformations for the classical and noise fields. We can see that the hydrodyanmic variables in~\eqref{eqn:hydro} are the currents for the transformations of the noise fields. 

Furthermore, the action must satisfy the EFT symmetries~\cite{liu2018lectures},
\begin{align}
I^*[\phi, \Phi, \psi_i, \Psi_i] &= -I[\phi, -\Phi, \psi_i, -\Psi_i],\nonumber\\
I[\phi, \Phi=0, \psi_i, \Psi_i=0] &=0, \nonumber\\
\text{Im}\, I[\phi, \Phi, \psi_i, \Psi_i] &\ge 0, \label{eqn:eftsym}
\end{align}
which can be derived from the full Schwinger-Keldysh formalism. It also must satisfy the KMS symmetry
\begin{align}
\phi(x,t) &\goesto -\phi(x,-t), \nonumber\\
\Phi(x,t) &\goesto -\Phi(x,-t) - i\beta\partial_t \phi(x,-t), \nonumber\\
\psi_i(x,t) &\goesto -\psi_i(x,-t), \nonumber\\
\Psi_i(x,t) &\goesto -\Psi_i(x,-t) - i\beta \partial_t \psi_i(x,-t), \label{eqn:KMS}
\end{align}
which is a consequence of the fact that our hydrodynamic EFT describes relaxation toward an equilibrium thermal thermal state $e^{-\beta H}$; similar ideas hold for more general steady state \cite{Guo:2022ixk}.

Lastly, we have the option of enforcing the ``diagonal shift symmetries" \cite{liu2018lectures}. These symmetries require that the action only depend on $\phi$ through $\partial_t \phi$, or only depend on $\psi_i$ through $\partial_t \psi_i$. In ordinary fluids, the EFT in the presence of the diagonal shift symmetry describes the normal phase, while the EFT in the absence of the diagonal shift symmetry describes superfluidity. Thus, condensed degrees of freedom need not obey the diagonal shift symmetry, while normal degrees of freedom must.

\invisiblesection{Hydrodynamic phases}
\textit{Hydrodynamic phases.}---
We will approach the hydrodynamics by imposing the diagonal shift symmetries for each phase independently, and then finding the lowest-order action in that phase. To count scaling dimensions, we note that $\psi_i$ must scale as $\grad_i\phi$ in order to preserve the dipole symmetry. First, we will impose the diagonal shift symmetry on both $\phi$ and $\psi_i$. This should describe the normal phase, with no condensation. If we suppose that the dynamical scaling exponent is $z=4$, the most general effective action consistent with the KMS and EFT symmetries is
\begin{align}
\L_\eff &= \chi \partial_t \phi \partial_t \Phi  + \big[ -\sigma \partial_t(\grad_i \phi - \psi_i)\big] (\grad_i \Phi - \Psi_i)  \nonumber\\
&\quad - \big[ B_1\partial_t \grad_i\grad_j \phi  + B_2 \partial_t \grad_i \psi_j + B_3 \partial_t \grad_j \psi_i \big] \grad_j \Psi_i,
\end{align}
to leading order. All coefficients must be positive, by a combination of the KMS and EFT symmetries and thermodynamic stability. We have named $\chi$ and $\sigma$ in reference to ordinary systems. Although $\chi$ is the susceptibility, $\sigma$ does not play the role of a measurable electrical conductivity.

The density and currents are, at leading order:
\begin{align}
\rho &= \chi\partial_t \phi \nonumber\\
\rho_i &= 0\nonumber\\
J_i &= - \sigma \grad_i \partial_t\phi + \sigma \partial_t\psi_i \nonumber\\
J_{ij} &= B_1 \grad_i\grad_j \partial_t \phi + B_2 \grad_i \partial_t \psi_j + B_3 \grad_j \partial_t \psi_i.
\end{align}
Although $\rho_i$ has nonzero contributions at higher order, we will not need to include them. The dipole continuity equation reads
\begin{align}
0 &= - \sigma \grad_i \partial_t\phi + \sigma \partial_t\psi_i - B_1 \grad^2\grad_i \partial_t \phi  \nonumber\\
&\quad - B_2 \grad_i\grad_j \partial_t \psi_j - B_3 \grad^2 \partial_t \psi_i
\end{align}
which imposes that $\partial_t\psi_i = \grad_i \partial_t\phi$, plus higher-order corrections. The monopole continuity equation then reads
\begin{align}
0 &= \partial_t \rho -\partial_t \grad_i \rho_i + \grad_i\grad_j J_{ij} \nonumber\\
&= \chi\partial_t^2 \phi + (B_1 + B_2 + B_3)\grad^4 \partial_t \phi + \cdots,
\end{align}
so that the dispersion is
\begin{align}
\omega = -i \frac{B_1 + B_2 + B_3}{\chi}k^4,
\end{align}
which describes subdiffusion. This is consistent with previous results~\cite{fractonhydro}, and also with experiments on cold atomic gases with approximate dipole symmetry~\cite{GuardadoSanchez2020}. 

For the remaining phases we will presciently suppose $z=2$. Then, the most general effective action consistent with the KMS and EFT symmetries, but without any diagonal shift symmetries imposed, is
\begin{align}
\L_\eff &= \chi \partial_t \phi \partial_t \Phi \nonumber\\
&\quad + \big[ -\kappa^\phi_1 (\grad_i \phi - \psi_i) + \kappa^\phi_2\grad^2\grad_i\phi +g_2 \grad^2\psi_i \nonumber\\
&\qquad +g_3 \grad_i\grad_j \psi_j - \sigma \partial_t(\grad_i \phi - \psi_i)\big] (\grad_i \Phi - \Psi_i)  \nonumber\\
&\quad - \big[ \kappa^{\phi\psi}\grad_i\grad_j \phi + \kappa^\psi_1 \grad_i \psi_j + \kappa^\psi_2 \grad_j \psi_i  \big] \grad_j \Psi_i,
\end{align}
to leading order. The $\kappa$ coefficients act as generalized superfluid stiffnesses in the system. The symmetries require that all coefficients except $g_2$, and $g_3$ are positive. Furthermore, $\kappa^{\phi\psi}=\kappa^\phi_2+g_2+g_3$ by KMS (see Appendix). This action should describe the charge condensate. Under these conditions, all terms in the effective action are allowed and the density and currents are
\begin{align}
\rho &= \chi\partial_t \phi \nonumber\\
\rho_i &= 0 \nonumber\\
J_i &= -\kappa^\phi_1\grad_i \phi + \kappa^\phi_1 \psi_i + \cdots \nonumber\\
J_{ij} &= \kappa^{\phi\psi} \grad_i\grad_j \phi + \kappa^\psi_1 \grad_i \psi_j + \kappa^\psi_2 \grad_j \psi_i,
\end{align}
to leading order. The dipole continuity equation now imposes that $\psi_i = \grad_i \phi$ plus higher-order corrections. The monopole continuity equation is 
\begin{align}
0 &= \partial_t \rho - \partial_t\grad_i\rho_i + \grad_i\grad_j J_{ij} \nonumber\\
&= \chi\partial_t^2 \phi + (\kappa^{\phi\psi} + \kappa^\psi_1 + \kappa^\psi_2)\grad^4 \phi + \cdots,
\end{align}
so that the dispersion is
\begin{align}
\omega^2 = \frac{\kappa^{\phi\psi} + \kappa^\psi_1 + \kappa^\psi_2}{\chi}k^4,
\end{align}
which describes a propagating mode with $\omega\sim k^2$. Going beyond leading order, including generic dissipative terms such as $ \partial_t \grad_i \psi_j \grad_i \Psi_j$ in the action contributes a subleading $-i k^4$ to the dispersion. 

The quadratic propagation matches previous expectations at $T=0$ from a microscopic model~\cite{Yuan2020}, field theory~\cite{Stahl2022}, and a more generic model called the Dipolar Bose-Hubbard Model (DBHM)~\cite{Lake2022}, so that the charge condensate behaves like a zero-temperature fluid. The effects of dissipation are subleading and do not modify the zero-temperature behavior at low wavevector. In Ref.~\cite{Lake2022}, the authors show that the existence of only a single mode in the charge condensate phase of the DBHM is a result of a Higgs-like effect. The same effect appears in the hydrodynamics as the requirement that $\psi_i = \grad_i\phi$.

Finally, we can try imposing the diagonal shift symmetry on $\phi$ but not $\psi_i$. This corresponds to the dipole condensate, where dipole symmetry is spontaneously broken but monopole symmetry is not. The diagonal shift symmetry on $\phi$ requires that $\kappa^\phi_1$, $\kappa^\phi_2$, and $\kappa^{\phi\psi}$ vanish, which in turn requires that $g_3=-g_2$.
The density and currents are
\begin{align}
\rho &= \chi\partial_t \phi \nonumber\\
\rho_i &= 0 \nonumber\\
J_i &= g_2 (\grad^2\psi_i - \grad_i\grad_j\psi_j) - \sigma \grad_i \partial_t\phi + \sigma \partial_t\psi_i \nonumber\\
J_{ij} &= \kappa^\psi_1 \grad_i \psi_j + \kappa^\psi_2 \grad_j  \psi_i. \label{eqn:dc-currents}
\end{align}
The dipole continuity equation will no longer result in a constraint because now $J_i$ and $\grad_jJ_{ij}$ are of the same order. Instead, we will have to simultaneously solve both equations. 

The two continuity equations are 
\begin{align}
0 &= \chi \partial_t^2 \phi - \sigma \grad^2 \partial_t \phi + \sigma \partial_t \grad_i\psi_i,\nonumber\\
0 &= -\sigma \grad_i \partial_t \phi + \sigma\partial_t \psi_i - (\kappa^\psi_2 - g_2) \grad^2 \psi_i \nonumber\\
&\hspace{2cm}- (\kappa^\psi_1 +g_2) \grad_i\grad_j \psi_j.
\end{align}
We can simplify the analysis by splitting $\psi_i$ into a transverse and longitudinal part $\psi_i = \psi_i^t + \psi_i^\ell$ where the longitudinal part is $\psi_i^\ell = k_ik_j/k^2 \psi_j$ and obeys $\grad_i\psi_i^\ell = \grad_i\psi_i$. The transverse part is $\psi_i^t = P_t\psi_j$ where $P_t = (\delta_{ij} - k_ik_j/k^2)$ is the transverse projector. Applying the transverse projector to the dipole continuity equation results in 
\begin{align}
0 &= \sigma \partial_t \psi_i^t - (\kappa^\psi_2 - g_2) \grad^2 \psi_i^t,
\end{align}
with solution
\begin{align}
\omega = -i \frac{\kappa^\psi_2-g_2}{\sigma} k^2, \label{eqn:dctran}
\end{align}
which is an ordinary diffusive mode. Note that the value $\kappa^\psi_2-g_2$ is always positive (see Appendix). Furthermore, this dispersion represents two hydrodynamic modes, corresponding to the two transverse polarizations of $\psi_i$.

To access the longitudinal part we may take the divergence of the dipole continuity equation. The monopole continuity equation and the divergence of the dipole continuity equation together read
\begin{align}
0 = 
\begin{bmatrix} \chi \partial_t - \sigma \grad^2 & \sigma \partial_t  \\
-\sigma \grad^2  & \sigma\partial_t - \kappa^\psi \grad^2
\end{bmatrix} \begin{pmatrix} \partial_t \phi \\ \grad_j\psi_j \end{pmatrix}, \label{eqn:dcmatrix}
\end{align}
where $\kappa^{\psi} = \kappa^\psi_1+ \kappa^\psi_2$, showing that $\phi$ and $\grad_i\psi_i$ are coupled. Their joint dispersion relation is
\begin{align}
0 = \omega^2 + i\frac{\kappa^{\psi}}{\sigma} \omega k^2 - \frac{\kappa^{\psi}}{\chi}k^4, \label{eqn:DCdisp}
\end{align}
or
\begin{align}
\omega = -i\frac{\kappa^{\psi}}{2\sigma}k^2 \pm \sqrt{\frac{-(\kappa^{\psi})^2}{4\sigma^2} + \frac{\kappa^{\psi}}{\chi}}k^2,
\end{align}
which displays a crossover from pure diffusion to quadratic propagation, controlled by the dimensionless parameter $\kappa^{\psi}\chi/\sigma^2$. For $\kappa^{\psi}\chi \gg 4\sigma^2$, the dispersion approaches
\begin{align}
\omega = -i\frac{\kappa^{\psi}}{\sigma} k^2, \qquad \omega = -i \frac{\sigma}{\chi} k^2,
\end{align}
with two quadratically diffusing modes. In the opposite limit the dispersion approaches
\begin{align}
\omega = -i \frac{\kappa^{\psi}}{2\sigma} k^2 \pm \sqrt{\frac{\kappa^{\psi}}{\chi}} k^2,
\end{align}
which is simultaneously quadratically propagating and quadratically diffusive. While we might have expected the dissipative coefficient $\sigma$ to play a damping role, the large-$\sigma$ regime is underdamped and the small-$\sigma$ regime is overdamped.

\invisiblesection{Exploring the dipole condensate}
\textit{Exploring the dipole condensate.}---
Since the subdiffusion of the normal phase and quadratic propagation of the charge condensate already exist in the literature, we can focus on understanding the dipole condensate better. We can tune various parameters to be small, bringing us to limiting points of the phase diagram. The small parameters define a \emph{quasihydrodynamic} timescale $\tau$~\cite{GrozdanovQuasiHydro}, which is parametrically long.

In particular, let us study the hydrodynamics in the charge condensate but near the transition to the dipole condensate. We allow terms that break the diagonal shift symmetry for $\phi$, but require them to be small. This defines the quasihydrodynamic timescale $\tau = \sigma/\kappa^\phi_1$ (see Appendix for more details). Furthermore, we must choose the dimensionless parameter $\kappa^{\psi}\chi/\sigma^2$ to place us on either side of the crossover in the dipole condensate. 
The resulting dispersion relation is shown in Fig.~\ref{fig:CCtoDC}. For details, see the Appendix.

\begin{figure}
    \centering
    \begin{tikzpicture}
    \draw (0, 0) node[inner sep=0] {\includegraphics[width=.45\columnwidth]{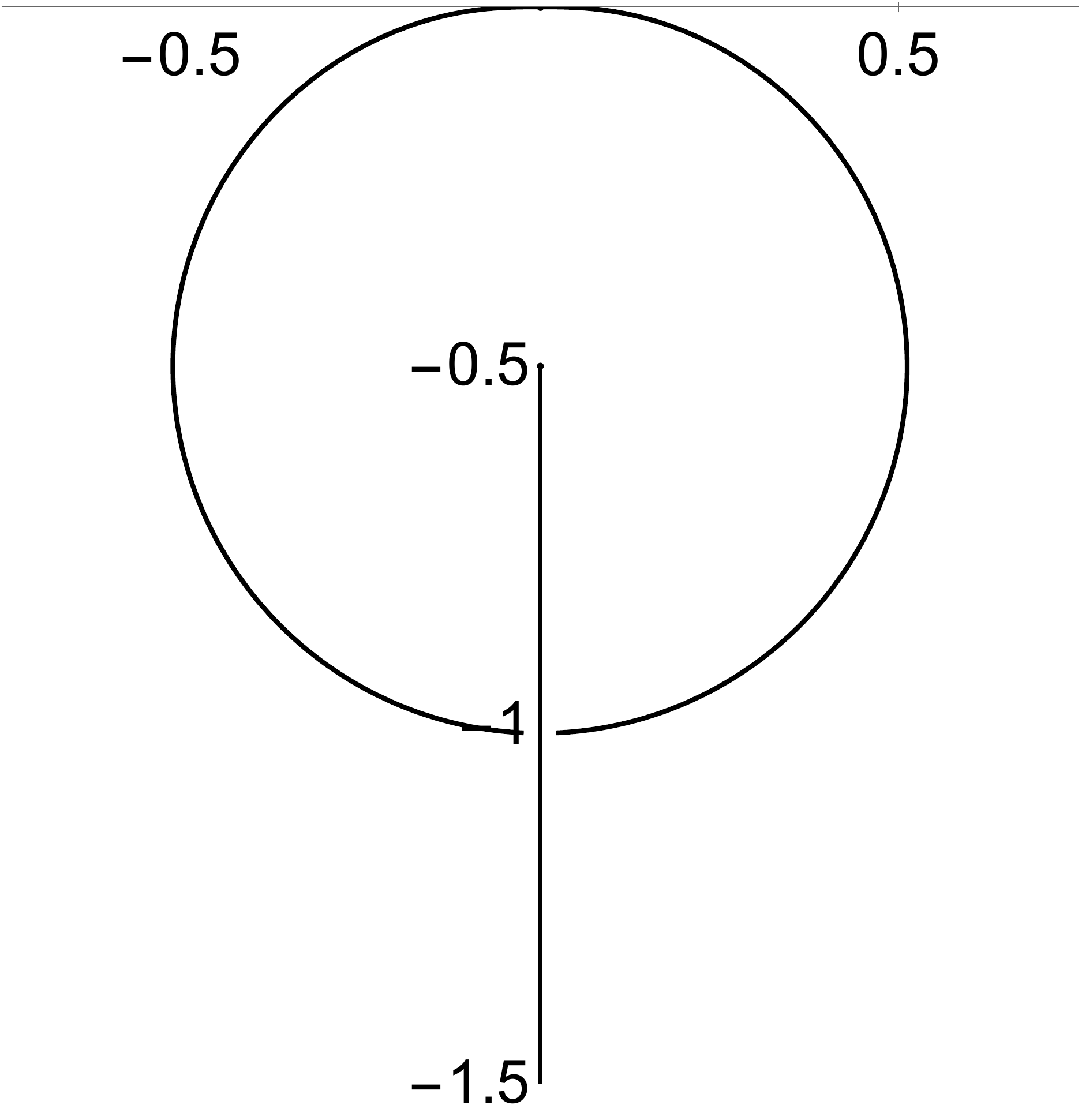}};
    \draw (2,1.8) node {Re $\omega$};
    \draw (.4,-1.4) node {Im $\omega$};
    \draw (0,2.4) node {$\begin{aligned}\frac{\kappa^\psi \chi}{\sigma^2}=25\end{aligned}$};
    \draw (4.5, 0) node[inner sep=0] {\includegraphics[width=.45\columnwidth]{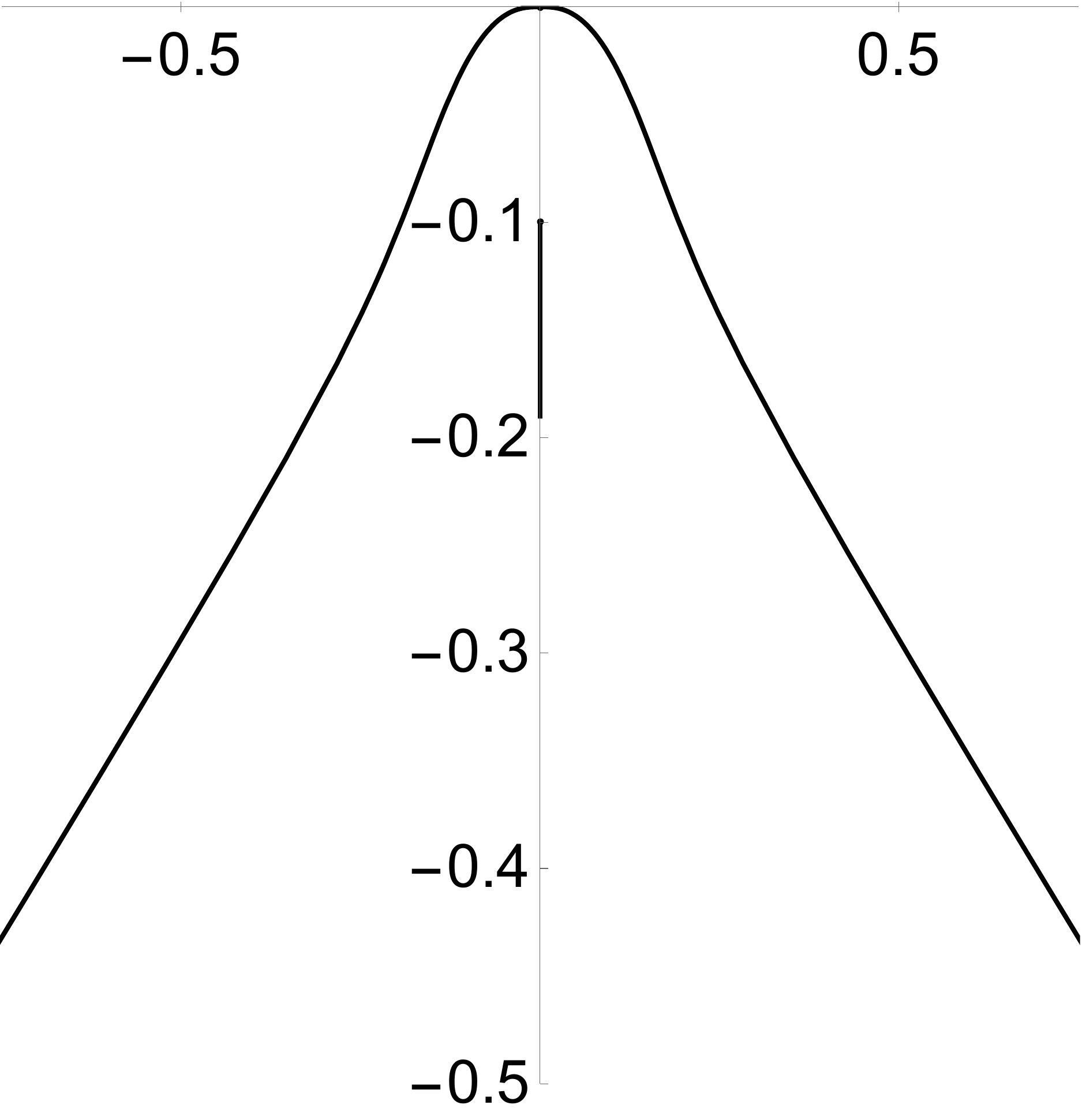}};
    \draw (6.5,1.8) node {Re $\omega$};
    \draw (4.9,-1.4) node {Im $\omega$};
    \draw (4.5,2.4) node {$\begin{aligned}\frac{\kappa^\psi \chi}{\sigma^2}=1\end{aligned}$};
    \end{tikzpicture}
    \caption{Parametric plot of the dispersion in the charge condensate but close to the dipole condensate. The left figure is plotted near the diffusive regime of the dipole condensate ($\kappa^{\psi}\chi/\sigma^2=25$) while the right figure is plotted near the propagating regime of the dipole condensate ($\kappa^{\psi}\chi/\sigma^2=1$). At small $k$ both dispersions look like the charge condensate while at large $k$ they look like their respective DC dispersions.}
    \label{fig:CCtoDC}
\end{figure}
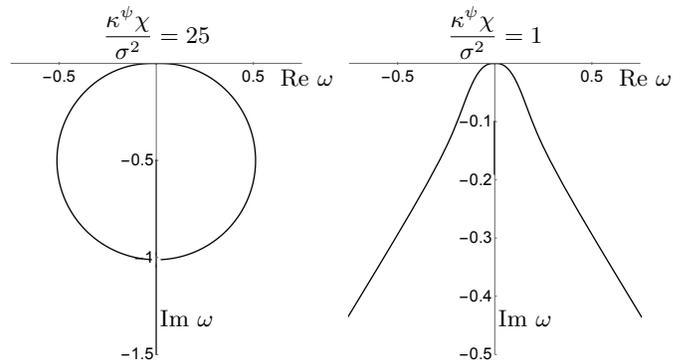

Another surprising facet of the phase is its quadratic propagation. At $T=0$ in the DBHM, the dipole condensate consists of $d$ modes (one for each space dimension), all propagating linearly~\cite{Stahl2022,Lake2022}. We can treat the hydrodynamic phase explored here as consisting of both the dipole condensate and a background normal (subdiffusive) fluid. Although the hydrodynamics EFT does not provide a mechanism for studying the behavior of the fluid as $T \goesto 0$, we can instead see that we can reproduce the $T=0$ behavior in the nondissipative limit $\sigma \goesto 0$. The dispersion at small $\sigma$ is in Fig.~\ref{fig:NoDiss} (see Appendix for details). In contrast to the charge condensate, where dissipation had little effect on the physics, in the dipole condensate the mode propagation at low wavenumber is immediately modified in the presence of dissipation. 

\begin{figure}
    \centering
    \begin{tikzpicture}
    \draw (0, 0) node[inner sep=0] {\includegraphics[width=.45\columnwidth]{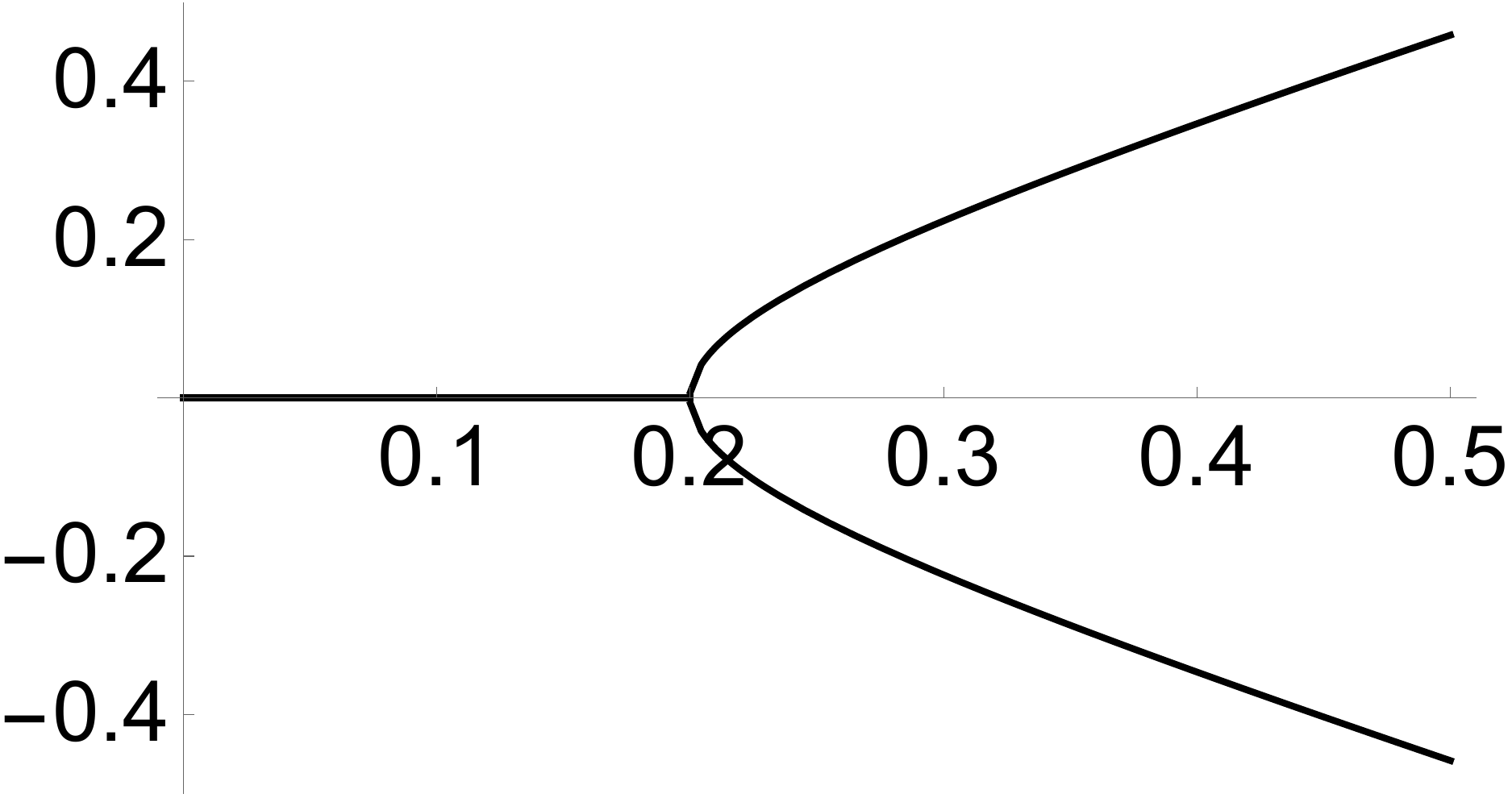}};
    \draw (2,.2) node {$k$};
    \draw (-1,.4) node {Re $\omega$};
    \draw (0,1.4) node {Re $\omega$};
    \draw (4.5, 0) node[inner sep=0] {\includegraphics[width=.45\columnwidth]{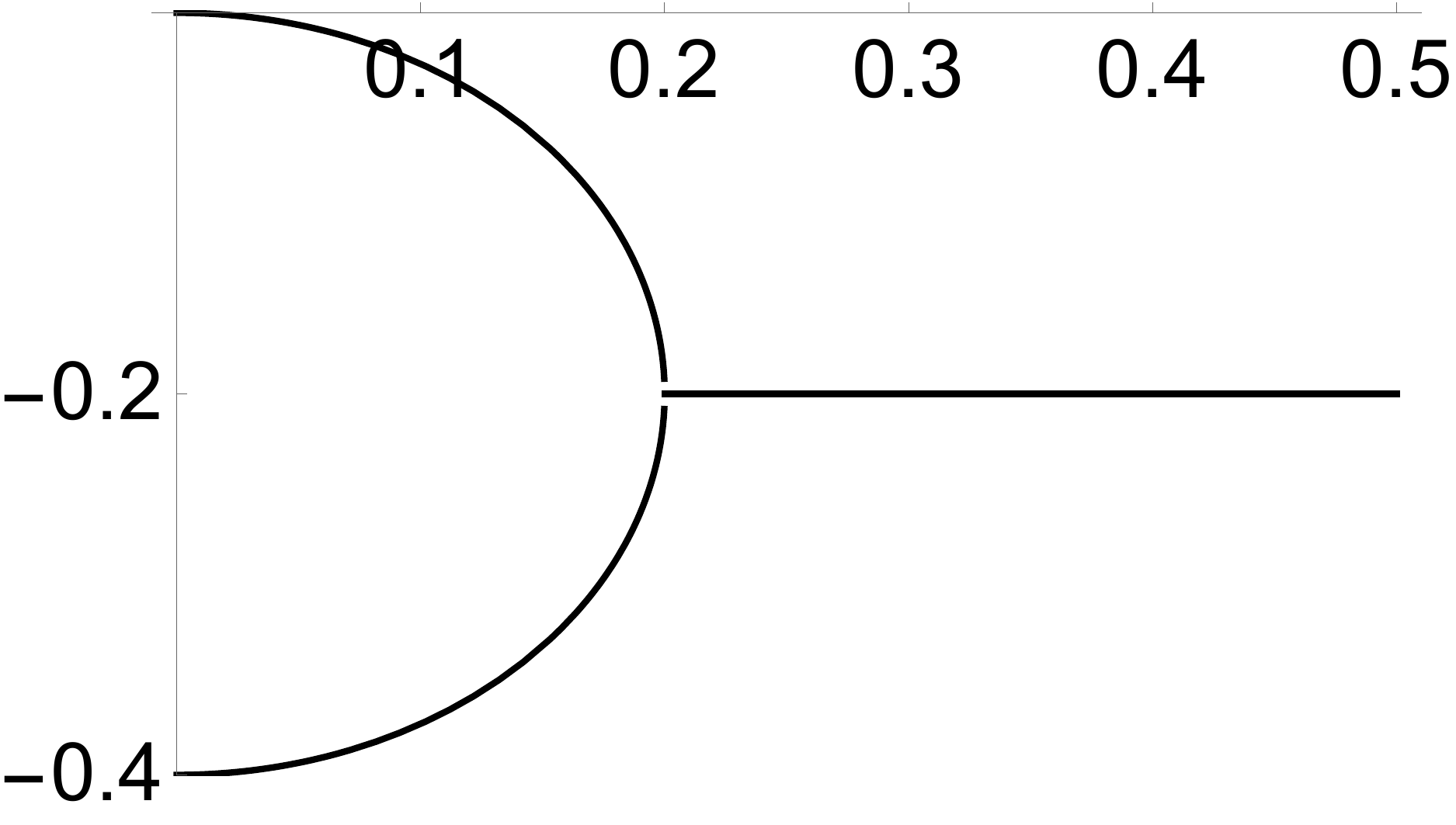}};
    \draw (6.5,1.2) node {$k$};
    \draw (3.5,.4) node {Im $\omega$};
    \draw (4.5,1.4) node {Im $\omega$};
    \end{tikzpicture}
    \caption{Real and imaginary parts of the dispersion in the dipole condensate phase close to the dissipationless limit. At small $k$ the dispersion looks like the diffusive regime of the dipole condensate $(\omega \sim -ik^2)$ while at large $k$ it looks like the $T=0$ limit of the dipole condensate $(\omega\sim\pm k)$.}
    \label{fig:NoDiss}
\end{figure}
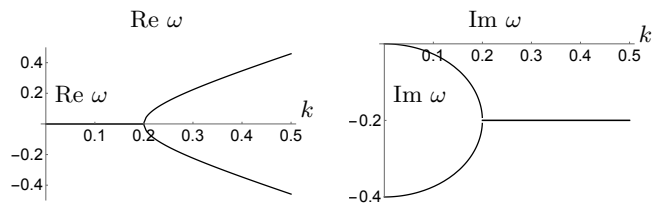

\invisiblesection{Discussion}
\textit{Discussion.}---
We have developed a systematic, effective field theory based treatment of hydrodynamics in systems with charge and dipole symmetry, allowing for the possibility of spontaneous symmetry breaking. In the absence of any SSB, we find quartic subdiffusion, consistent with \cite{fractonhydro}. With both charge and dipole symmetries broken, we find a quadratically propagating (and quartically subdiffusing) mode, consistent with \cite{Lake2022}. We also introduced the phase where dipole symmetry is spontaneously broken but monopole symmetry is preserved, corresponding to a `dipole condensate.' In this phase we find that there exist diffusive transverse modes, as well as longitudinal modes which depending on parameters can be either purely diffusive, or quadratically propagating and relaxing. This phase does not match any in the literature, and reflects intrinsically-nonzero-temperature effects.

Our results can be tested in ultracold atom experiments analogous to \cite{GuardadoSanchez2020}. Further afield, they could be generalized to systems with momentum conservation and/or systems with more complex multipolar symmetries \cite{BHN}. We leave such generalizations to future work. 

\invisiblesection{Acknowledgements}
\textit{Acknowledgements.} CS and RN were supported by the U.S. Department of Energy, Office of Science, Basic Energy Sciences, under Award DE-SC0021346. MQ was supported by U.S. Department of Energy, Office of Science, Basic Energy Sciences under Award number DE-SC0014415. AL was supported by NSF CAREER Grant DMR-2145544, and by the Alfred P. Sloan Foundation under Grant FG-2020-13795. PG was supported by the Department of Energy through Award DE-SC0019380, the Simons Foundation through Award No. 620869, and the Alfred P. Sloan Foundation under Grant FG-2020-13615.

\invisiblesection{Appendix: Quasihydrodynamics}
\textit{Appendix: Quasihydrodynamics.}---
To study the quiasihydrodynamics mentioned in the main text, we need to choose a parameter to tune to be small. For the transition from the charge condensate into the dipole condensate,  we will reintroduce the coefficients $\kappa^\phi_1$ and $\kappa^{\phi\psi}$ ($\kappa^\phi_2$, $g_2$, and $g_3$ will all be subleading). This gives a longitudinal continuity equation,
\begin{align}
0 = \omega^2 + i\frac{\kappa^{\psi}}{\kappa^\phi_1-i\omega \sigma}\omega^2 k^2 - \frac{\kappa^{\phi\psi}+\kappa^\psi}{\chi}k^4, \label{eqn:CCtoDC}
\end{align}
with $\kappa^{\psi}=\kappa^\psi_1+\kappa^\psi_2$ as before. This dispersion defines a time scale $\tau = \sigma/\kappa^\phi_1$. After this time, the dispersion is $\omega^2 = (\kappa^{\phi\psi}+\kappa^\psi) k^4 / \chi$, reproducing the charge condensate. This shows that we are truly in the charge condensate. Before this scale, the dispersion looks like the dipole condensate (compare to~\eqref{eqn:DCdisp}). There is no transverse mode in the charge condensate, even near the dipole condensate transition.

We can better understand~\eqref{eqn:CCtoDC} by looking at plots of the dispersion. In Fig.~\ref{fig:CCtoDC} we can see the dispersion with $\kappa^{\psi}\chi/\sigma^2=25$ chosen to place us firmly within the diffusive regime.  At small $k$ (true hydrodynamics) the dispersion looks like the charge condensate with $\omega^2 \sim k^4$, with an additional gapped mode.
At large $k$ (quasihydrodynamics) it looks like the diffusive regime of the dipole condensate $(\omega\sim -ik^2)$.  One of the diffusive modes becomes gapped at small $k$, while the other diffusive mode collides with the large-$k$ gapped mode to give the propagating modes. 

Fig.~\ref{fig:CCtoDC} also shows the dispersions at a value of $\kappa^{\psi}\chi/\sigma^2=1$ (recall the critical value is 4). The propagating modes at small $k$ become the propagating modes at large $k$, with no collision. There is an extra mode that is gapped at large and small $k$. This mode is not a hydrodynamic or quasihydrodynamic mode, but it cannot be removed from the analysis because it is the same mode that goes from diffusive to gapped in the other regime.

To study the small-$T$ regime of the dipole condensate, let us revisit~\eqref{eqn:dc-currents} in the small-$\sigma$ limit. This allows us to retain $\rho_i$ in the continuity equations, with the important contribution being $\rho_i = \chi^\psi \partial_t\psi_i$. The transverse part of the dipole continuity equation reads
\begin{align}
0 = \chi^\psi \partial_t^2 \psi_i^t + \sigma \partial_t \psi_i^t - (\kappa^\psi_2-g_2) \grad^2 \psi_i^t,
\end{align}
with solution
\begin{align}
\omega = \frac{-i\sigma}{2\chi^\psi} \pm \sqrt{\frac{-\sigma^2}{2(\chi^\psi)^2} + \frac{(\kappa_2^\psi - g_2)k^2}{\chi^\psi}},
\end{align}
introducing a timescale $\tau = \chi^\psi/\sigma$. The new timescale $\tau$ is large in the small-$\sigma$ limit. On timescales smaller than $\tau$ the quasihydrodynamics consists of a linear propagating mode,
\begin{align}
\omega = \pm \sqrt{\frac{(\kappa_2^\psi - g_2)k^2}{\chi^\psi}},
\end{align}
matching the $T=0$ expectation. At timescales larger than $\tau$ the propagating mode splits into a gapped mode and a diffusive mode with diffusion constant $(\kappa_2^\psi - g_2)/\sigma$, as in~\eqref{eqn:dctran}.

With the introduction of $\chi^\psi$, the analog of~\ref{eqn:dcmatrix} is 
\begin{align}
0 = 
\begin{bmatrix} \chi \partial_t - \sigma \grad^2 & \sigma \partial_t  \\
-\sigma \grad^2  & \chi^\psi \partial_t + \sigma\partial_t - \kappa^\psi \grad^2
\end{bmatrix} \begin{pmatrix} \partial_t \phi \\ \grad_j\psi_j \end{pmatrix},
\end{align}
with the same timescale $\tau = \chi^\psi/\sigma$. In the small-$\sigma$ limit, the solutions are 
\begin{align}
\omega = -i\frac{\sigma}{\chi}k^2,\qquad \omega  = \frac{-i\sigma}{2\chi^\psi} \pm \sqrt{\frac{-\sigma^2}{2(\chi^\psi)^2} + \frac{\kappa^\psi\,k^2}{\chi^\psi}}.
\end{align}
The first solution matches one of the diffusion modes from the dipole condensate phase, with a diffusion constant that vanishes in the small-$\sigma$ limit. The other mode behaves like the transverse mode, transitioning from linear propagation in quasihydrodynamics to a gapped mode and a diffusive mode in the late-time hydrodynamics. These modes are shown in Fig.~\ref{fig:NoDiss}.

The above analysis shows that the dissipative coefficient $\sigma$ is crucial in that it completely changes the nature of the dispersion relation from $T=0$ to finite $T$, going from a ballistic to a quadratic scaling. While we determined the presence of this transport coefficient in terms of simple symmetry arguments, we note that this term can be argued to be finite based on microscopic reasoning. Consider a lattice model described by a complex boson $b_{\bm x}$ and with dipole symmetry $b_{\bm x}\to b_{\bm x} e^{i {\bm \alpha}\cdot {\bm x}}$. In the condensed dipole phase, hopping of a single boson is allowed through the term $b_{\bm x}b_{\bm x+\bm e_j}e^{i\psi_j}+\text{ h.c.}$, where $\bm e_j$ denotes a unit vector in the $j$-direction \cite{Lake2022}. Note that $\psi_i$ can exactly be viewed as a spatial gauge field $A_i=\psi_i$. Treating $\psi_i$ as a backround non-dynamical field, at finite temperature, this coupling will generically lead to a finite conductivity term in the current $J_i=\sigma E_i=\sigma\partial_t \psi_i$, which is precisely the last term in the third line of~\eqref{eqn:dc-currents}. This argument not only confirms that $\sigma$ \emph{must} generically be finite, it also shows that, given a $U(1)$-invariant system without dipole symmetry, this can be straightforwardly extended to a dipole symmetric system in the dipole condensed phase.

\invisiblesection{Appendix: Derivation of the effective action}
\textit{Appendix: Derivation of the effective action.}---
The effective actions we consider must obey the KMS symmetry in~\eqref{eqn:KMS}. Ref.~\cite{Kapustin2022} shows that we can construct KMS-invariant terms in two distinct ways, which correspond to dissipative and nondissipative terms in the effective action. The nondissipative terms are
\begin{align}
\L_\text{nd} &= \left( \Phi \frac{\delta}{\delta \phi} + \Psi_i \frac{\delta}{\delta \psi_i} \right) \int d^3x\, dt\, \Omega,
\end{align}
where $\Omega$ is a Lagrangian that depends on $\phi$ and $\psi_i$ but not on $\Phi$ or $\Psi_i$. Thermodynamic stability of the effective action requires that $\Omega$ is negative when Wick-rotated. The dissipative terms are
\begin{widetext}
\begin{align}
\L_\text{d} &= \frac{1}{2} \left( X(\phi,\psi_i,\Phi,\Psi_i) + X_\text{KMS}(\phi,\psi_i,\Phi,\Psi_i) - X(\phi,\psi_i,0,0) - X_\text{KMS}(\phi,\psi_i,0,0) \right),
\end{align}
where $X$ is quadratic in $\Phi$ and $\Psi_i$ and is even under time-reversal. The function $X_\text{KMS}$ is the result of the transformation in~\eqref{eqn:KMS} applied to $X$.

For the nondissipative part, we will consider terms of order $\omega^2$, $\omega^2k^2$, $k^2$, and $k^4$. This is not a strictly valid gradient expansion at any value of $z$, but will give us all the terms we need for our analysis. Then, we have
\begin{align}
2\Omega &= \chi (\partial_t \phi)^2 + \chi_2^\phi (\partial_t\grad_i\phi)^2 + 2g_1 \partial_t\grad_i\phi \partial_t \psi_i +\chi^\psi (\partial_t\psi_i)^2 - \kappa^\phi_1(\grad_i\phi-\psi_i)^2 \nonumber\\
&\quad - \kappa^\phi_2(\grad_i\grad_j\phi)^2 - 2g_2 \grad_i\grad_j \phi \grad_i\psi_j - 2g_3 \grad^2\phi \grad_i\psi_i - \tilde{\kappa}_2^\psi( \grad_i\psi_j)^2 - \tilde{\kappa}_1^\psi(\grad_i\psi_i)^2,
\end{align}
where we have included various factors of 2 for convenience. All $\chi$ and $\kappa$ coefficients must be nonnegative. The $g$ coefficients may be positive or negative, but must obey the stability conditions $|g_1| \le \min(\chi_2^\phi,\chi^\psi)$, $|g_2| \le \min(\kappa^\phi_2,\tilde{\kappa}_2^\psi)$, $|g_3| \le \min(\kappa^\phi_2,\tilde{\kappa}_1^\psi)$, and $|g_2+g_3| \le \kappa^\phi_2$. The Lagrangian becomes
\begin{align}
\L_\text{nd} &= \left[\chi \partial_t\phi - \chi_2^\phi\partial_t\grad^2\phi - g_1\partial_t\grad_i\psi_i \right] \partial_t\Phi \nonumber\\
&\quad + \left[ g_1\partial_t\grad_i\phi + \chi^\psi\partial_t\psi_i \right] \partial_t \Psi_i \nonumber\\
&\quad + \left[ -\kappa^\phi_1(\grad_i\phi-\psi_i) \right] (\grad_i\Phi - \Psi_i)\nonumber\\
&\quad + \left[ -\kappa^\phi_2\grad_i\grad_j\phi - g_2\grad_i\psi_j  - g_3\delta_{ij}\grad_k\psi_k\right] \grad_i\grad_j\Phi \nonumber\\
&\quad + \left[ - g_2\grad_i\grad_j\phi - g_3 \delta_{ij}\grad^2\phi - \tilde{\kappa}_2^\psi \grad_i\psi_j - \tilde{\kappa}_1^\psi\delta_{ij} \grad_k\psi_k \right] \grad_i\Psi_j \nonumber\\
&= \left[\chi \partial_t\phi - \chi_2^\phi\partial_t\grad^2\phi - g_1\partial_t\grad_i\psi_i \right] \partial_t\Phi \nonumber\\
&\quad + \left[ g_1\partial_t\grad_i\phi + \chi^\psi\partial_t\psi_i \right] \partial_t \Psi_i \nonumber\\
&\quad + \left[ -\kappa^\phi_1(\grad_i\phi-\psi_i) + \kappa^\phi_2 \grad^2\grad_i\phi + g_2\grad^2\psi_i + g_3\grad_i\grad_j\psi_j \right] (\grad_i\Phi - \Psi_i)\nonumber\\
&\quad - \left[ (\kappa^\phi_2+g_2+g_3) \grad_i\grad_j\phi + (\tilde{\kappa}_1^\psi+g_3) \grad_i\psi_j + (\tilde{\kappa}_2^\psi+g_2)\grad_j\psi_i \right] \grad_j\Psi_i,
\end{align}
where we used $\grad_i\grad_j\Phi = \grad_i(\grad_j\Phi - \Psi_j)+\grad_i\Psi_j$ and integration by parts. Note the sign and order of indices in the last line, chosen to match the convention in~\eqref{eqn:hydro}. We can identify the new coefficients   $\kappa^{\phi\psi} = \kappa^\phi_2+g_2+g_3$, $\kappa_1^\psi = \tilde{\kappa}_1^\psi+g_3$, and $\kappa^\psi_2 = \tilde{\kappa}_2^\psi+g_2$, all of which are nonnegative.

The dissipative terms we need for our analysis descend from the expression
\begin{align}
2\beta X = i b_0(\grad_i\Phi - \Psi)^2 + i b_1(\grad_i\grad_j\Phi)^2 + 2i\xi \grad_i\grad_j\Phi \grad_i\Psi_j + 2i\xi_2 \grad^2\Phi \grad_i\Psi_i + ib_2 (\grad_i\Psi_j)^2 + ib_3(\grad_i\Psi_i)^2,
\end{align}
where the $b$ coefficients must be positive and $|\xi_1| \le \min(b_1,b_2)$, $|\xi_2| \le \min(b_1,b_3)$, and $|\xi_1+\xi_2| \le b_1$ by~\eqref{eqn:eftsym}. Then,
\begin{align}
\L_\text{d} &= X - b_0\partial_t(\grad_i\phi - \psi_i)(\grad_i\Phi - \Psi_i) \nonumber\\
&\quad - \left[ b_1 \partial_t \grad_i\grad_j\phi + \xi_1\partial_t \grad_i\psi_j + \xi_2 \delta_{ij}\partial_t\grad_k\psi_k \right] \grad_i\grad_j \Phi \nonumber\\
&\quad -\left[ \xi_1\partial_t\grad_i\grad_j\phi + \xi_2\delta_{ij} \partial_t \grad^2\phi + b_2\partial_t\grad_i\psi_j + b_3\delta_{ij} \partial_t\grad_k\psi_k \right] \grad_i\Psi_j\nonumber\\
&= X + \left[ - b_0\partial_t(\grad_i\phi - \psi_i) + b_1\partial_t\grad^2\grad_i\phi + \xi_1\partial_t\grad^2\psi_i + \xi_2\partial_t\grad_i\grad_j\psi_j \right](\grad_i\Phi - \Psi_i) \nonumber\\
&\quad -\left[ (b_1+\xi_1+\xi_2) \partial_t\grad_i\grad_j\phi + (b_2+\xi_1) \partial_t\grad_i\psi_j + (b_3+\xi_2) \delta_{ij} \partial_t\grad_k\psi_k \right] \grad_i\Psi_j,
\end{align}
\end{widetext}
from which we can identify $\sigma=b_0$, $B_1 = b_1+\xi_1+\xi_2$, $B_2 = b_2+\xi_1$, and $B_3 = b_3+\xi_2$. The other terms end up being sub-leading so we may drop them. The terms in $X$ itself are quadratic in $\Phi$ and $\Psi_i$, so they contribute to the fluctuating hydrodynamics but can be ignored for the purpose of computing the dispersion relations.

\bibliography{hydro}

\end{document}